\documentclass[useAMS,usenatbib,referee]{mnras}
\usepackage{amsmath,bm}
\usepackage{amssymb}

\usepackage{graphics}
\usepackage{epstopdf}
\usepackage{epsf}

\def\p{\partial}

\def\b{\boldsymbol}

\def\sun{\odot}
\newcommand{\lambdabar}{{\mkern0.75mu\mathchar '26\mkern -9.75mu\lambda}}
\usepackage{graphicx}
\usepackage{graphics}
\usepackage{epstopdf}
\usepackage{epsf}

\title{Magnetic absorption of VHE photons  in the magnetosphere of the Crab pulsar}

\author[S.V.Bogovalov et al.]
{	S. V.~Bogovalov$^{1}$,
	I.~Contopoulos$^{1,2}$, 
	A.~Prosekin\thanks{E-mail:prosekin@mpi-hd.mpg.de}$^{3}$, 
	I.~Tronin$^{1}$,
	F. A.~Aharonian$^{1,3,4}$
    \\
    $^{1}$National Research Nuclear University (MEPHI), Kashirskoje shosse, 31, Moscow, Russia\\
    $^{2}$Research Center for Astronomy and Applied Mathematics, Academy of Athens, Athens 11527, Greece\\
    $^{3}$Max-Planck-Institut f\"ur Kernphysik, Saupfercheckweg 1, 69117 Heidelberg, Germany\\
    $^{4}$Dublin Institute for Advanced Studies, School of Cosmic Physics, 31 Fitzwilliam Place, Dublin 2, Ireland\\}

\begin{document}

\date{}

\pagerange{\pageref{firstpage}--\pageref{lastpage}} 

\maketitle

\label{firstpage}

\begin{abstract}
The detection of the pulsed $\sim 1 $~TeV gamma-ray emission from the Crab pulsar reported by MAGIC and VERITAS collaborations demands a substantial revision of existing models of particle acceleration in the pulsar magnetosphere. In this regard model independent restrictions on the possible production site of the VHE photons become an important issue. In this paper, we consider limitations imposed by the process of conversion of VHE gamma rays into $e^{\pm}$ pairs in the magnetic field of the pulsar magnetosphere. Photons with energies exceeding 1~TeV are effectively absorbed even at large distances from the surface of the neutron star. Our calculations of magnetic absorption in the force-free magnetosphere show that the twisting of the magnetic field due to the pulsar rotation makes the magnetosphere more transparent compared to the dipole magnetosphere. The gamma-ray absorption appears stronger for photons emitted in the direction of rotation than in the opposite direction. There is a small angular cone inside which the magnetosphere is relatively transparent and photons  with energy $1.5$~TeV can escape from distances beyond $0.1$~light cylinder radius ($R_{\rm{lc}}$). The emission surface from where photons can be emitted in the observer's direction further restricts the sites of VHE gamma-ray production. For the observation angle $57^{\circ}$ relative to the Crab pulsar axis of rotation and the orthogonal rotation, the emission surface in the open field line region is located as close as $0.4\,R_{\rm{lc}}$ from the stellar surface for a dipole magnetic field, and $0.1\,R_{\rm{lc}}$ for a force-free magnetic field.  

\end{abstract}

\begin{keywords}
pulsars, gamma rays
\end{keywords}

\section{Introduction}
The detection of  a pulsed component of very high energy  (VHE) gamma rays  up to $\sim 1.5$~TeV from the Crab pulsar \citep{Aleksic2012,VERITASCollaboration2011, Ansoldi2016} demands a  serious revision of the current models of gamma-ray production in the pulsar magnetospheres. In our previous studies we have proposed \citep{Bogovalov2000}  and later elaborated \citep{Aharonian2012} the scenario where  VHE gamma-rays are produced beyond the light cylinder via IC scattering of the electrons in the ultrarelativistic pulsar wind. { It was shown that in such kind of scenario the absorption of VHE gamma-rays is negligible.} In this paper, we examine whether the more standard scenario, namely gamma-ray production inside of the light cylinder (hereafter the pulsar magnetosphere), can explain the observations of pulsed TeV gamma-ray emission.  In general, the hypothesis of magnetospheric  origin of TeV gamma-rays faces two major problems. Firstly, the dominant gamma-ray production mechanism in pulsars, namely {\it curvature radiation}, can hardly provide an extension of the energy spectrum beyond 0.1 TeV. Secondly, even assuming  that somehow the TeV gamma-rays are produced, their escape from the magnetosphere is hampered  due to pair-production in the strong radiation and magnetic fields. Here we do not consider the first problem, but simply postulate gamma-ray production and focus on the study of the transparency of the pulsar magnetosphere for TeV gamma-rays.

In accordance with standard pulsar models, electrons in the pulsar magnetosphere are accelerated in the electrostatic gaps formed in charge depleted regions. Several gap models have been proposed.
The so-called polar cap models \citep{Daugherty1996} are based on the assumption that the gap is located right above the polar cap. Its thickness is of a few hundred meters. Magnetic pair creation $\gamma B\rightarrow e^{\pm}$ and pair cascades predict a super-exponential cutoff in the region $\sim 10$~GeV due to the sharp dependence of the pair production probability on the photon energy. On the other hand, super-exponential  spectral turnovers in the Fermi gamma-ray data are
ruled out to a high degree of significance \citep{Abdo2009,Abdo2010}. Therefore, it is quite unlikely that the polar cap region is the site of VHE gamma-ray production, at least within the framework of the initial "standard" model. Later, this model evolved into the slot gap model which extends much higher in altitude \citep{Muslimov2004}. A different model with a gap extending from the null charge surface to the light cylinder (the so-called outer gap) has been extensively studied by many authors. This model predicts a simple exponential cutoff in the energy spectrum \citep{Chiang1994}. Nevertheless, all gap models have a serious difficulty to account for the VHE pulsed radiation observed from the Crab. Acceleration in the electrostatic gap is accompanied by the formation of counterstreaming fluxes of charged particles and photons in a rather wide energy band. In the case of Crab, this radiation is detected from visible light to soft X-rays. The interaction of accelerated electrons with the soft non-thermal radiation results in the formation of electromagnetic cascades in which the observed gamma-ray spectra are formed. However, the same radiation field limits the Lorentz factor of the primary electrons to a value of a few times $10^5$ \citep{shibata} and prevents the leakage of VHE gamma-rays from the magnetosphere. In the case of the Crab pulsar, the mean free path of photons in relation to the $\gamma\gamma$ interaction appears less than a $0.01$ fraction of the light cylinder radius.

To overcome this problem, one may assume that the radiating electrons are accelerated by a different mechanism which does not produce counterstream fluxes of electrons and photons. Possible alternatives are magnetocentrifugal acceleration or acceleration in the turbulent plasma \citep{Machabeli1979,Blandford1982,bog14,Osmanov2009}. In this case, both electrons and positrons move in the same direction producing phase coincident fluxes of photons in all energy bands observed from the Crab pulsar. As a result, the $\gamma\gamma$ pair production is suppressed and VHE photons can freely escape from the magnetosphere. 

While the transparency of the magnetosphere in relation to $\gamma\gamma$ interaction strongly depends on model of the particle acceleration,  the magnetic absorption of gamma-rays is almost model-independent. It depends only on the geometry and strength of the magnetic field which has been obtained by several independent research groups \citep[e.g.][etc.]{Spitkovsky2006,Kalapotharakos2009,Tchekhovskoy2016, Petri2016}.
The opacity of the pulsar magnetosphere in relation to the $\gamma B$ conversion of photons in pairs has been discussed in the past in the context of electromagnetic cascades in the polar cap where the main cascade mechanism is pair production due to this process. \citep{Daugherty1982,Arons1979,Hibschman2001}. 
Recently, this process has been revisited, motivated by new gamma-ray observations in the GeV and TeV bands.
Nevertheless, the opacity of the pulsar magnetosphere in relation to photons with energy of the order of $\sim 1$~TeV was not investigated in detail taking into account all applicable mechanisms.  In the recent paper by \citep{Story2014} the calculations were performed for the dipole magnetic field. This is a reasonable approach for energies below $10$~GeV because gamma-rays of these energies are absorbed close to the pulsar surface where the distortion of the magnetic field due to rotation can be neglected. For photons with energies close to 1 TeV  the absorption occurs at higher altitudes where this distortion cannot be ignored.
The magnetic field of the rotating pulsar is twisted in the azimuthal direction and provides additional components compared to the initial dipole configuration. In the magnetic field gamma-rays are emitted almost along field lines. On the other hand, the conversion of a photon into a pair occurs when the photon reaches a sufficient pitch angle due to the curvature of the magnetic field lines. Thus, one can expect that the increase of the curvature of field lines due to their rotational twisting should result in the increase of VHE photon opacity of the magnetosphere. However, as our calculations show, the twisting of field lines in the direction opposite to pulsar rotation leads to the reduction of opacity.

In the present work we derive the general limitations on the potential sites of production of high energy gamma rays imposed by the mere existence of the rotating magnetosphere with the force-free condition. These include the surface (hereafter the $\tau=1$ surface) separating the region of gamma-ray emission with strong magnetic absorption from the region with free escape without absorption, the surface emitting photons in the observer's direction (the emission surface), and the surface of last closed field lines restricting the possible emission volume to the region of open field lines. The photons are emitted in the direction of motion of relativistic electrons, i.e. at some angle to the magnetic field lines due to electric drift. This angle can also be treated as the result of the aberration appearing in the transformation from locally co-rotating frame. The similar technique has been used by \cite{Contopoulos2010,Bai2010,Bai2010a} to define the phase curve of radiation below $10$~GeV. The rotation defines not only the emission direction through aberration, but also influences the absorption probability during photon propagation through rotating field lines. 

The positions of the limiting surfaces depend on the rotation, the observation angle (for the emission surface), the strength and structure of the magnetic field. The latter has been obtained earlier in two approximations, MHD \citep{Tchekhovskoy2013} and force free \citep{Contopoulos1999,Kalapotharakos2009,Timokhin2006,Petri2016}. In this paper we use the force-free magnetosphere of the orthogonal rotator calculated in the manner similar to previous calculations performed by \cite{Kalapotharakos2012}. In a realistic pulsar magnetosphere it is necessary to take into account the electrostatic gaps where the $e^{\pm}$ plasma is produced and accelerated. However, in this work we will assume that the force-free approximation holds everywhere. For comparison, we also present the results obtained for the dipole approximation of the pulsar magnetosphere.

The rest of the paper is organized as follows. Section~\ref{sec:2} discusses the strength and the structure of magnetic field along with other relevant pulsar parameters. Section~\ref{sec:3} presents the results of calculations and explores their general features.  The discussion and conclusions are given in Section~\ref{sec:4}. The basic steps of our calculations are described in Appendices~\ref{sec:aconv} and \ref{sec:acalc}.

\section{Magnetosphere parameters}
\label{sec:2}

\subsection{Estimate of the magnetic field}
The magnetic field of a solitary pulsar can be estimated assuming that the decrease of the rotational energy is caused by magneto-dipole radiation with the intensity determined by the equation
\begin{equation}
L_{\rm dipole}= {B_p^2\Omega^4 R_{\ast}^6\over 6c^3}\sin^2{\alpha},
\end{equation}
where $B_p$ is the magnetic field at the pole of the neutron star, $\Omega$ is the angular velocity, $R_{\ast}$ is the radius of the  neuron star, and $\alpha$ is the inclination angle. This estimate, however,  contains a large uncertainty. Pulsars eject plasma in the form of a wind. The density of the pulsar wind is sufficient to screen the magneto-dipole electromagnetic radiation. Therefore, pulsars lose rotational energy due to a magnetized wind which carries out the angular momentum and energy of the pulsar rotation. A more accurate equation for the rotational losses has been proposed by \cite{Spitkovsky2006} in the following form
\begin{equation}
L_{\rm{rot}}\approx \frac{3}{2}L_{90^\circ {\rm dipole}}(1+\sin^2{\alpha})\ ,
\label{spit}
\end{equation}
where $L_{90^\circ {\rm dipole}}\equiv B_p^2\Omega^4 R_{\ast}^6/6c^3$.
For the orthogonal rotator these losses are three times higher than the losses due to the magneto-dipole radiation.

The rotational energy losses result in the increase of the period of rotation $P$ of the pulsar with the rate $\dot P$ defined from observations. Following to the conventional procedure \citep{Manchester1978} and using Eq.~(\ref{spit}), the magnetic field at the pole of the neutron star can be estimated as
\begin{equation}
B_p=4.7\times 10^{19}(1+\sin^2{\alpha})^{-1/2} (P\dot P)^{1/2} \left({R_{\ast} \over 10^6 {\rm cm}}\right)^{-2} \left({M\over M_{\sun}}\right)^{1/2} \rm G,
\end{equation}
where $M$ is the mass of the neutron star.

The largest uncertainties in pulsar magnetic field estimates arise from uncertainties in the mass and radius of the neutron star. From observations of binary systems containing pulsars we know that the masses  lie in the range $1.3 - 2 M_{\sun}$. These masses are estimated with an accuracy better than $1\% $ \citep{Antoniadis2013,Demorest2010,Kramer2008}.The radii of neutron stars derived from observations of their thermal radiation lie in a rather wide range of $10-20$~km \citep{Potekhin2014}. On the other hand, equations of state place the radius of the neutron star closer to $15$~km  \citep{Haensel2007}. In this work we will assume a neutron star mass $M=1.5\,M_{\sun}$ and radius $R_{\ast}=15~ \rm km$. In this case the dipole magnetic field at the poles equals $B_p=2.1\times 10^{12}$~G. This is the value we use in our calculations. Note that the dipole magnetic field depends on the radius as $B_{\rm dip}\sim B_pR^{-3}$. It implies that our field corresponds to the value $7\times 10^{12}$~G at the distance $10$ ~km. As the position of the $\tau=1$ surface scales as $r_{\tau=1}\sim B_{\rm dip}^{\alpha}$ with $\alpha=0.25-0.5$ (depending on direction), the results depend relatively weakly on the mass and star radius. The calculations show that in the force-free magnetosphere this dependence is even weaker.

\subsection{Magnetosphere in the force-free approximation}

The structure of a realistic pulsar magnetosphere is a key issue for calculations of the optical depth. Meanwhile, numerical simulations of the pulsar magnetosphere in different approximations have significantly deepened our understanding of this still not fully resolved problem \citep{Contopoulos1999,Timokhin2006,Komissarov2006,McKinney2006,Yu2011,Parfrey2012,Cao2016a,Kalapotharakos2009,Spitkovsky2006}. These studies show that the magnetic field around and beyond the light cylinder is dramatically distorted by rotation.
The distortion of the dipole magnetic field may be attributed to the ``inertia'' of the electromagnetic field \citep{Bogovalov2001}. Calculations show that in young pulsars like Crab, the flux of electromagnetic field energy exceeds the kinetic energy by several orders of magnitude. This means that the inertia of the particles of the plasma is small compared to the inertia of the electromagnetic field, thus it is natural to neglect it (force-free approximation).

The structure of the pulsar magnetosphere in the force-free approximation is characterized by the following equations (e.g., \cite{Gruzinov1999,Gruzinov2005})
\begin{eqnarray}
 {\p {\bf E}\over \p t}=c\nabla\times {\bf B} -4\pi {\bf J},\\\nonumber
 {\p {\bf B}\over \p t}=-c\nabla\times {\bf E},\qquad \\\nonumber
  \nabla\cdot \bf B=0, \hspace{2cm}
\end{eqnarray}
where the density of electric current in the magnetosphere is defined by the equation  
\begin{equation}
 {\bf J}=\rho_e c{\bf E\times B\over B^2}+{c\over 4\pi}{(\bf B\cdot \nabla\times B-E\cdot\nabla\times E)\over B^2} \bf B,
\end{equation}
while 
\begin{equation}
 \rho_e={1\over 4\pi} \nabla\cdot E
\end{equation}
is the electric charge density. The numerical method for the calculation of the 3D force-free pulsar magnetosphere has been described in detail by \cite{Spitkovsky2006,Kalapotharakos2009}. 

The actual inclination angle of the Crab pulsar is not well known. Based on phenomenological radio pulse shape models \citep{Lyne2013} it is estimated to lie between $45^\circ$ and $75^\circ$. However, these estimates are model dependent. There are arguments  in favor of an inclination angle close to $90^\circ$ \citep{Komissarov2013}. In the present work, we restricted our consideration to the magnetosphere of an orthogonal rotator to simplify understanding of the  3D structure of the possible radiation sites.
  
\section{Results of calculations}
\label{sec:3}
We calculated the gamma-ray opacity of the dipole and the force-free magnetosphere, assuming orthogonal pulsar rotation.
In the calculations the coordinate axis $z$ is oriented along rotation axis of pulsar, whereas $x$-axis of the rotating reference frame is oriented along magnetic momentum of the star. Bearing in mind the results of MAGIC collaboration \citep{Ansoldi2016}, we consider $E_{\gamma}=10^{11},\, 4\times10^{11}, \, 1.5\times 10^{12}$~eV as the representative values of the photon energies.

\subsection{Transparency of the magnetosphere in dipole approximation}
We search for the emission region of photons which accumulate optical depth $\tau \le 1$ before they escape the magnetosphere. The emission positions with $\tau = 1$ after propagation form a surface constraining the region from where photons cannot escape. For the verification of our method, our calculations of the magnetic field opacity were performed both analytically and numerically
The optical depth of the magnetic field was calculated analytically using the method of steepest descent used by \cite{Arons1979} and exploited for calculations of the magnetic opacity by \cite{Hibschman2001} and \cite{Story2014}. The general steps of the numerical calculations are described in Appendices~\ref{sec:aconv} and \ref{sec:acalc}. In Fig.~\ref{fig1} the analytical and numerical calculations are compared. 
\begin{figure}
\centering
\includegraphics[width=\textwidth]{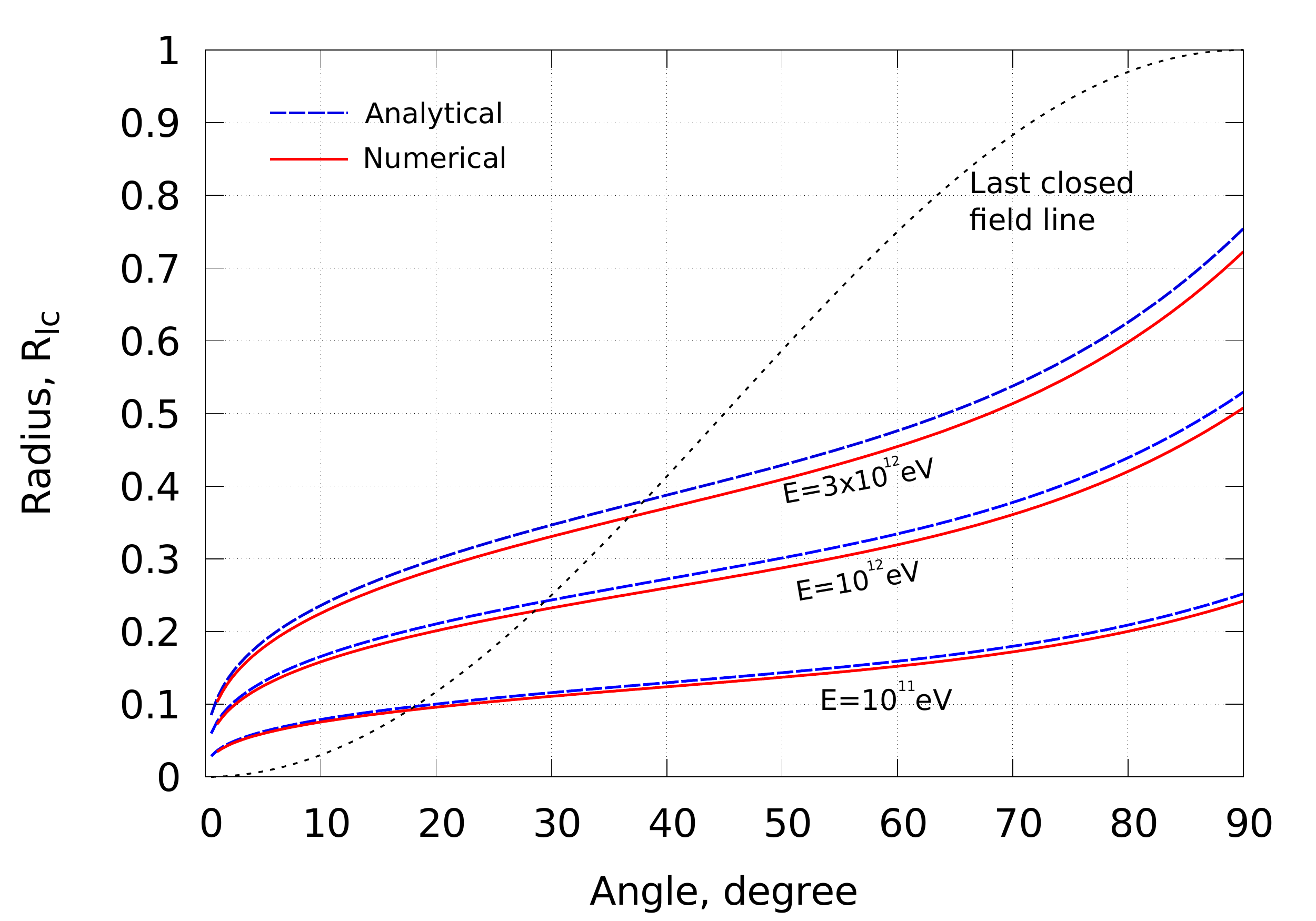}
 \caption{Verification of the calculation method. The radius of the $\tau=1$ surface vs. the angle on the equatorial plane for the dipole magnetic field with $B_p=4\times10^{12}$~G at the pole of the pulsar with $R_*=10$~km. The angle is measured from the magnetic dipole direction. Solid/dashed lines correspond to the numerical/analytical calculations, respectively.}
\label{fig1}
\end{figure}
One can see a good overall agreement between two calculations. Slight differences result from the approximations  used in the analytical approach.

\begin{figure}
\includegraphics[width=\textwidth]{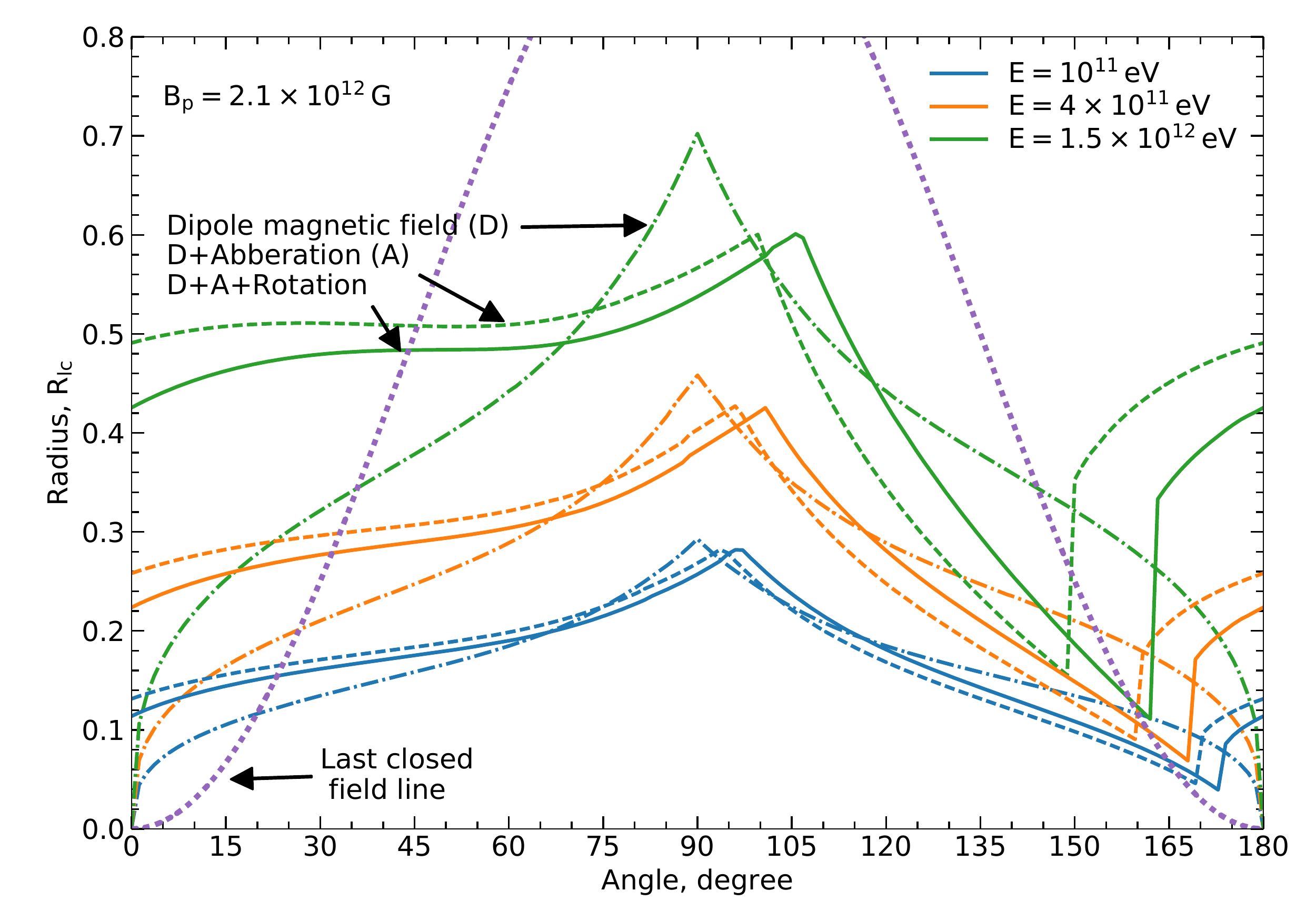}
 \caption{The radius of the $\tau=1$ surface vs. angle in the equatorial plane calculated for the orthogonal rotator under different assumptions and for different energies. The dash-dotted lines are for a non-rotating dipole field, the dashed lines are for calculations taking into account aberration, and the solid lines are for calculations taking into account the aberration and rotation of the magnetic field. The dotted line is the last closed field line of the dipole magnetic field. The angle is measured from the magnetic dipole direction in the direction of rotation (as shown in Fig.~\ref{fig:proj_z0}).}
\label{fig2}
\end{figure}

Fig.~\ref{fig2} shows the impact of aberration and rotation of the magnetic field on the gamma-ray opacity in the case of the dipole magnetic field. The effect of rotation makes the magnetosphere more opaque in the direction of rotation and more transparent in the opposite direction.  The aberration effect alone makes the surface more asymmetric, whereas the effect of rotation reduces this asymmetry. It is interesting to compare our results with the results of \cite{Story2014} predicting that on the last closed field line, the $\tau=1$ surface for $E_{\gamma}=4\times10^{11}$~eV is located at a distance of $0.2\,R_{\rm{lc}}$ from the star. In our case, this distance is found to be equal to $0.28\,R_{\rm{lc}}$ on the last closed field line along the direction of rotation, and $0.1\,R_{\rm{lc}}$ along the direction opposite to the direction of rotation. It is seen the value $0.2\,R_{\rm{lc}}$ is reached only when the rotation of the magnetosphere is neglected.

\subsection{Transparency of the force-free magnetosphere}

\begin{figure}
	\centering

	\mbox{\includegraphics[width=\textwidth]{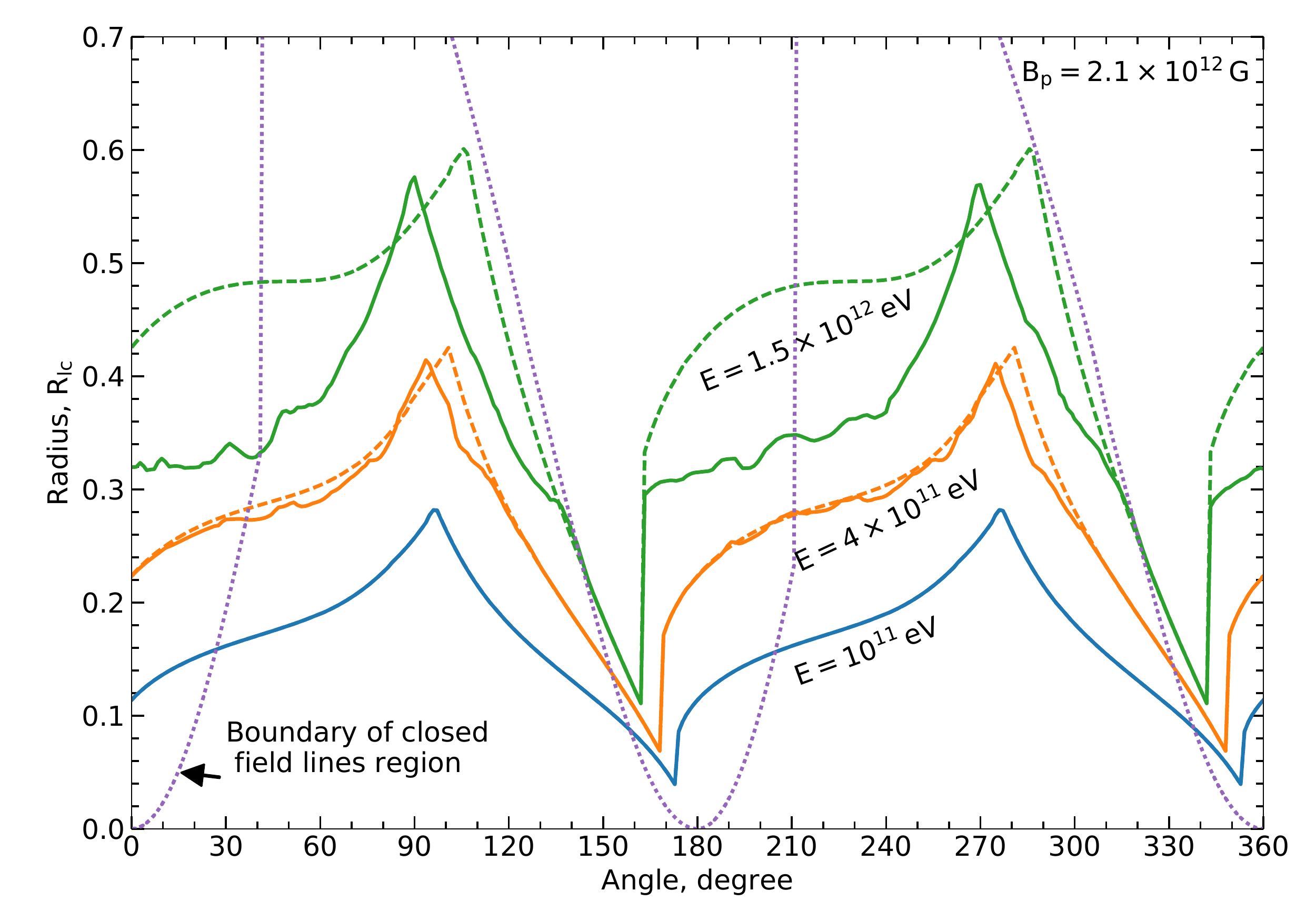}}
	
	\begin{minipage}[b]{\textwidth}
		\caption{The surfaces of optical depth $\tau=1$ in the XY plane at $z=0$ perpendicular to the rotation $z$-axis in radius-angle coordinates for different photon energies. The angle is measured from the magnetic dipole direction in the direction of rotation. The strength of the magnetic field at the pole is $B_p=2.1\times10^{12}$~G for a pulsar radius of $R_*=15$~km. The solid and dashed lines represent results for the force-free and dipole magnetic field, respectively. The dotted line presents the boundary of the region of closed field lines of the force-free magnetosphere in the XY plane.}\label{fig:tau1_rang}
	\end{minipage}

\end{figure}

\begin{figure}
	\centering

	\mbox{\includegraphics[width=0.5\textwidth]{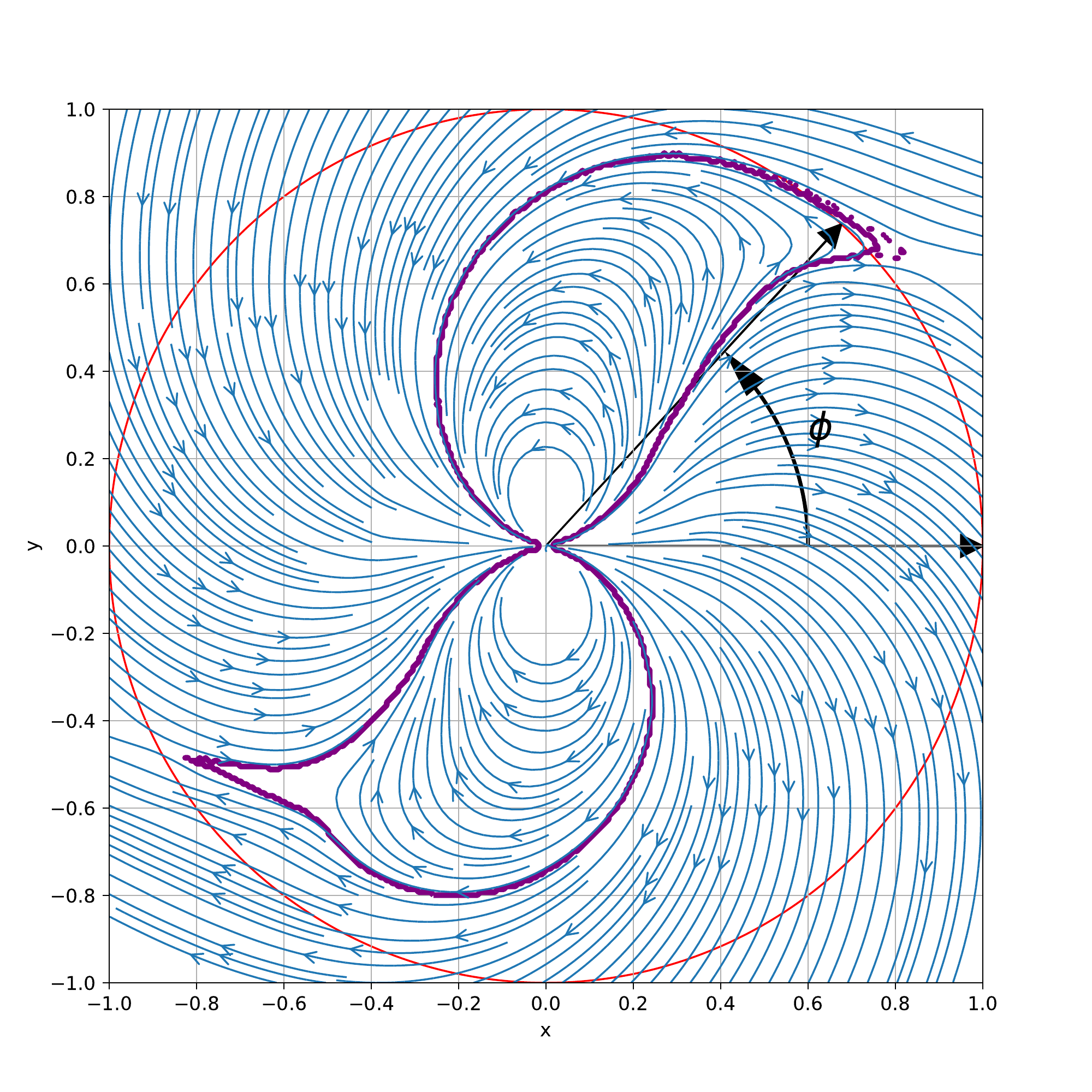}
		\includegraphics[width=0.5\textwidth]{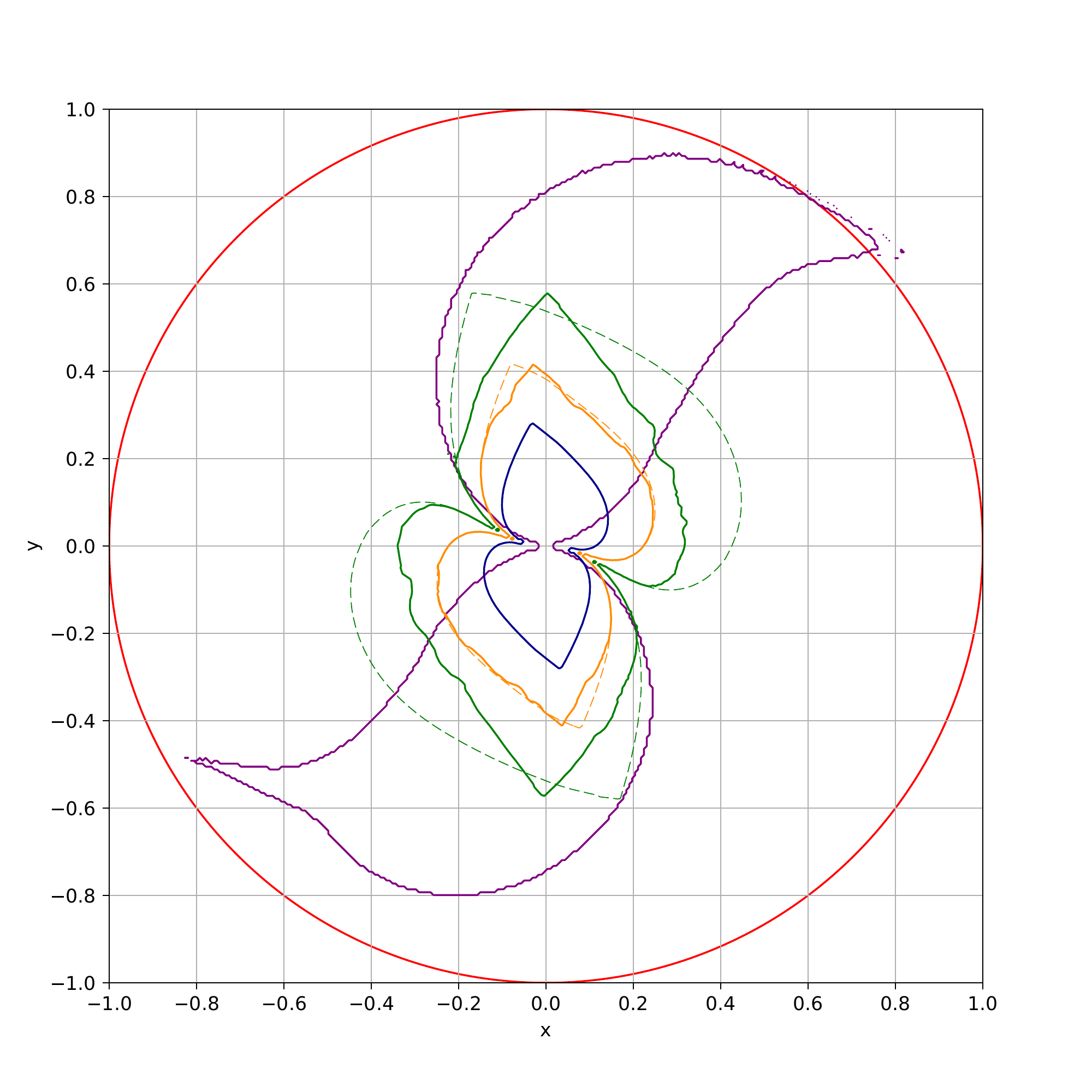}}
	
	\begin{minipage}[b]{\textwidth}
		\caption{Left panel: snapshot of equatorial field lines after 2 full pulsar rotations in a force-free magnetospheric calculation. The whole magnetosphere is 'breathing' (field lines that touch the light cylinder and the tip of the closed line region open and close dynamically), hence the asymmetry. Red circle: light cylinder. Violet line: boundary of the region of closed field lines. The direction of rotation is shown with an arrow. The angle $\phi$ in Fig.~\ref{fig:tau1_rang} is measured from the $x$-axis in the direction of rotation. Right panel: Fig.~\ref{fig:tau1_rang} in Cartesian coordinates.}\label{fig:proj_z0}
	\end{minipage}

\end{figure}

We performed a force-free electrodynamics (FFE) numerical calculation of an orthogonal rotating dipole in Cartesian coordinates. We followed the procedure described in detail in \cite{Spitkovsky2006,Kalapotharakos2009}. The calculation was performed in Cartesian coordinates centered on the star with a numerical grid resolution of $0.01\,R_{\rm{lc}}$. \mbox{At $t=0$} we start with a dipole magnetic field along the $x$-axis, and set the central star in rotation around the $z$-axis. The evolution of the simulation is very dynamic. The magnetosphere becomes severely distorted; the field lines that cross the light cylinder are stretched to larger and larger distances and become open. However, the numerical diffusion allows the field lines to close and open dynamically causing the tip of the closed line region to move in and out (but close to) the light cylinder. The steady state of our simulation is a continuous ``breathing'' of the magnetosphere. Contrary to previous FFE calculations (e.g. \cite{Kalapotharakos2012}), we did not implement any averaging, central symmetry, nor smoothing in our calculation, hence the asymmetry of the configuration shown in the equatorial plane after 2 full rotations in the left panel of Fig.~\ref{fig:proj_z0}. The magnetosphere is strongly distorted by the rotation of the pulsar in the vicinity of the light cylinder. The magnetic field is twisted in the direction opposite to the direction of rotation. As the "star" in the force-free simulations has the radius of $0.3\,R_{\rm{lc}}$ we join the force-free solution with the dipole magnetic field at $0.35\,R_{\rm{lc}}$ for the magnetosphere used in the calculations. The contours of the $\tau=1$ surface are shown in the right panel.

The dependence of the radius of the $\tau=1$ surface on the azimuthal angle $\varphi$ is shown in Fig.~\ref{fig:tau1_rang}. Here we compare the results for the force-free magnetosphere with the results for the dipole magnetic field. At energies below 400 GeV the curves in the dipole and force-free field coincide because the surface is located close to the pulsar. At energy $1.5$~TeV the difference starts to become remarkable. The irregular character of the curves for the force-free magnetosphere is related to scattering of the values of the magnetic field in the region of the current sheet. Note that at higher energies the $\tau=1$ surface appears to be located closer to the pulsar than for the dipole magnetic field. This means that the magnetic absorption of the VHE photons in the force-free magnetosphere is reduced compared to that for the dipole magnetic field.

The smallest absorption takes place at the poles. The shortest distances of the emission positions for which photons acquire $\tau=1$ in the magnetosphere are $r_{\rm{min}}=0.035\,$, $0.06\,$ and $0.1\,R_{\rm{lc}}$ at the angles $\phi_{\rm{min}}=174^{\circ}, 170^{\circ}, 163^{\circ}$ (and at the opposite to them angles $\phi_{\rm{min}}+180^{\circ}$) for photon energies $E_{\gamma}=1$, $4$ and $15\times 10^{11}$~eV, respectively. The change of the angles of closest positions leads to slight "rotation" of the $\tau=1$ surface at the poles with increase of the photon energy in the direction opposite to pulsar rotation. Fig.~\ref{fig:tau1_rang} demonstrates that, along the equatorial plane, photons moving along field lines twisted in the direction of rotation are absorbed essentially stronger than photons moving along field lines twisted in the opposite direction. The $\tau=1$ surface on the last closed field line in the direction of rotation is located at distances $0.16\,$, $0.27\,$, and $0.33\,R_{\rm{lc}}$ for photons with energies $1$, $4$, and $15\times10^{11}$~eV, respectively. The same surface in the opposite direction is located closer to the pulsar at distances $0.09\,$, $0.14\,$, and $0.25\,R_{\rm{lc}}$, respectively, as the absorption is suppressed there.

\begin{figure}. 
 \centering
   \mbox{\includegraphics[width=0.5\textwidth]{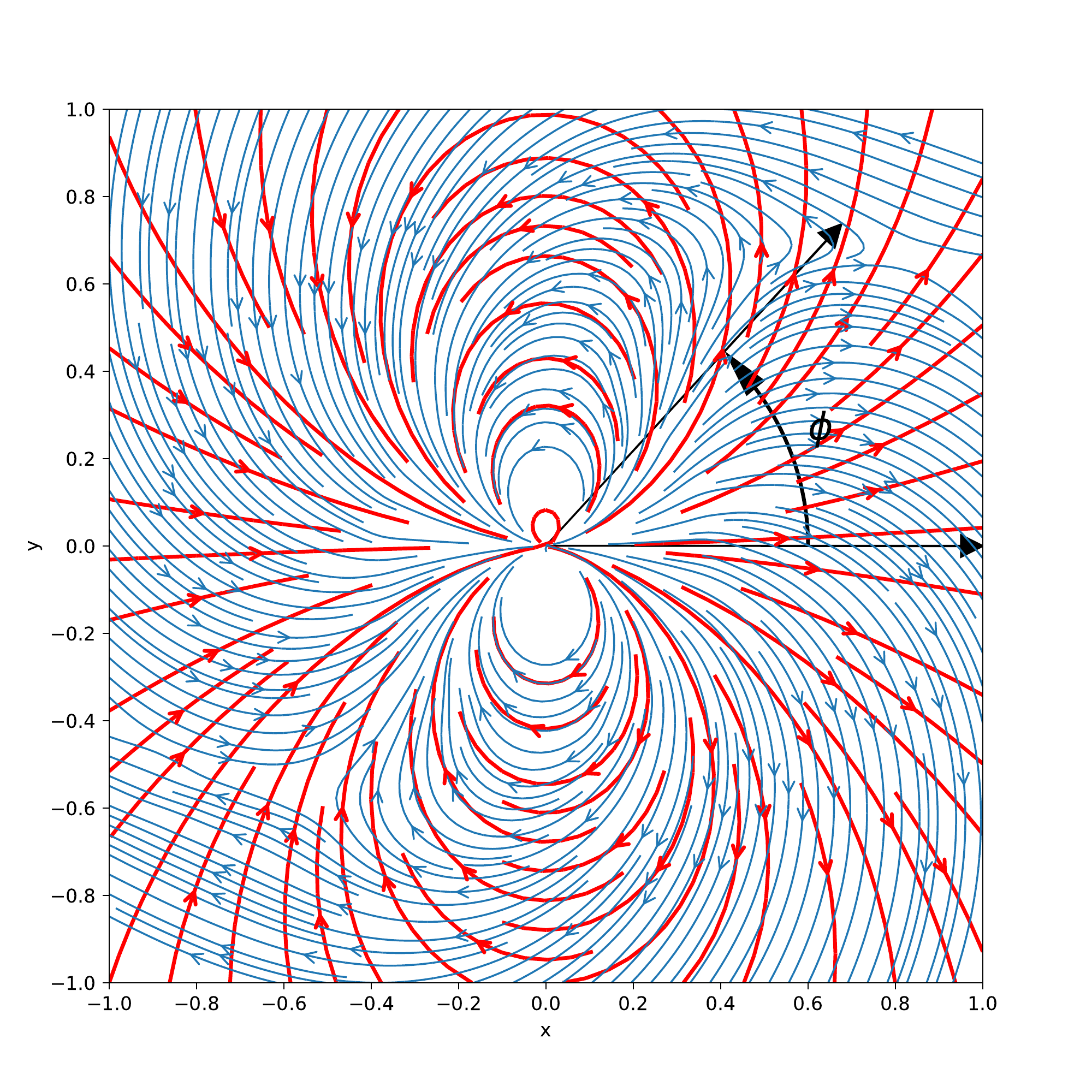}
   	\includegraphics[width=0.5\textwidth]{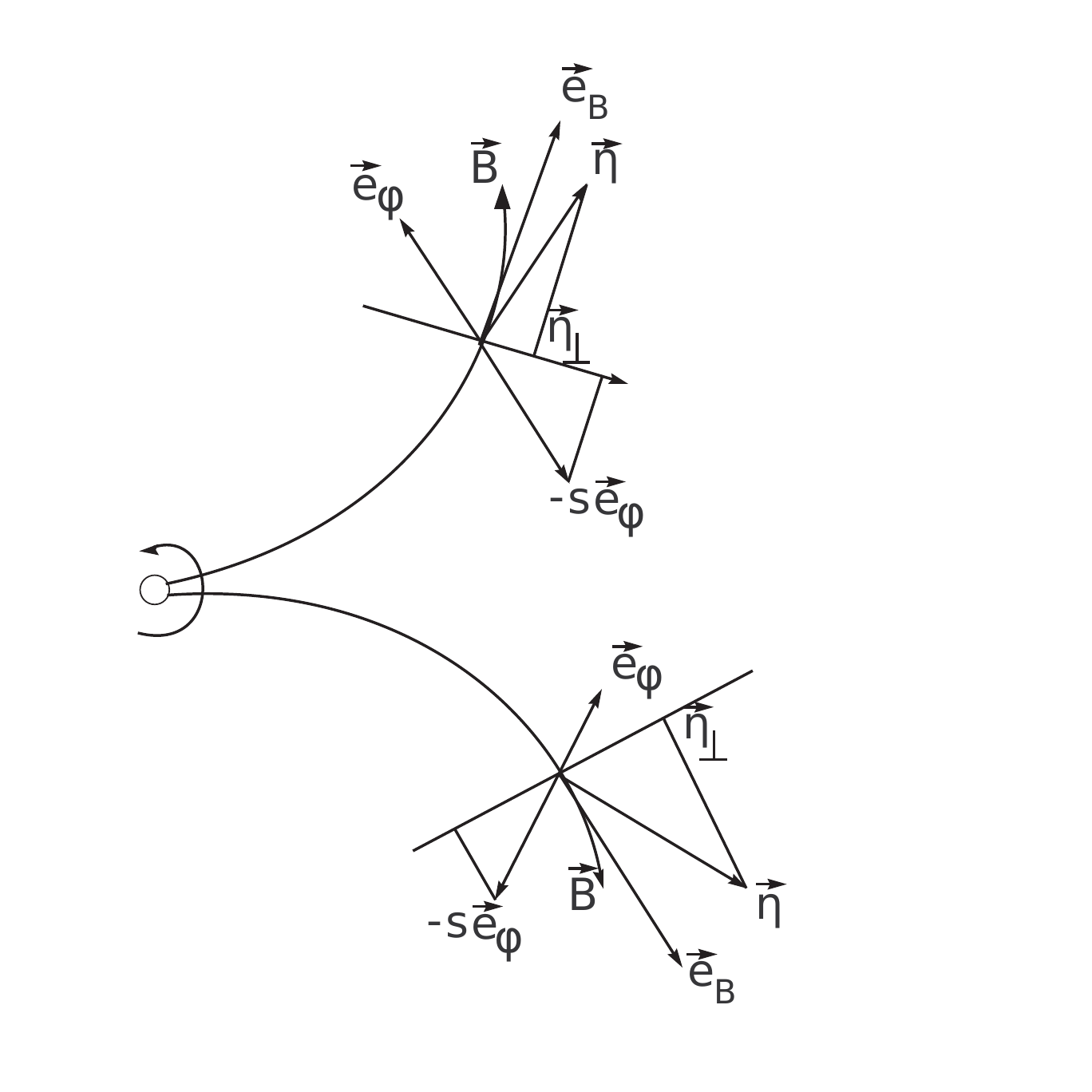}}
\caption{Left panel: comparison of force-free (blue) and dipole (red) magnetic field lines in equatorial plane. Right panel: the geometry of vectors $\b \eta,~\b e_{\phi},~\b e_{B}$ and $-s\b e_{\phi}$ on the magnetic field lines twisted in and opposite to the direction of rotation. Components of the vectors  $(\b\eta_{\perp}$ and $-s\b e_{\phi})$ perpendicular to $\b B$ are summed on the field lines twisted in the direction of rotation (here $\kappa'$ increases)  and are subtracted on the field lines twisted in the direction opposite to the direction of rotation (here $\kappa'$ decreases).}
\label{fig:vectors}
\end{figure}

{ One can notice that both the larger transparency of the force-free magnetosphere compared to the dipole magnetosphere and the asymmetry of the absorption relative to the magnetic moment are related to bending direction of field lines. The bending in the direction of rotation, as in the case of the dipole field for angles $\phi\lesssim 90^{\circ}$ in Fig.~\ref{fig:vectors}, makes magnetosphere more opaque, whereas the bending in the opposite direction, as for angles $\phi\gtrsim 90^{\circ}$, makes it more transparent. At the same time the open field lines of the force-free magnetosphere tend to twist in the direction opposite to the direction of rotation forming a spiral (see Fig.~\ref{fig:proj_z0}), which transforms into Archimedean spiral beyond light cylinder. This increases the transparency compared to the dipole magnetosphere. 

The influence of the field line bending on the absorption can be explained as follows. In the presence of the electric field $\b E\cdot\b B=0$, $E<B$ there is a reference frame moving with electric drift velocity $\b V=(\b E\times \b B)/B^2$ where the electric field is zero. The absorption rate per unit length in the observer's reference frame can be calculated using Eq.~(\ref{eq:att_length}) with $\kappa'=|\b B'\times \b k'|/(B_{cr}mc^2)$ calculated in this drift reference frame (see Appendix for details). Here $\b k=\epsilon \b \eta$, where $\epsilon$ and $\b \eta$ are the photon energy and direction. In the limit $\kappa\ll 1$
the absorption rate can be expressed as 
\begin{equation}\label{eq:attl1}
\mathcal{R}=0.23\alpha e B_{cr}\frac{\kappa'}{\epsilon}e^{-\frac{8}{3\kappa'}},
\end{equation}
from where it is obvious that absorption rate increases with $\kappa'$. Using Eq.~(\ref{eq:abber}) and $\b B'=\b B/\Gamma$ we can express $\kappa'$ through the values in the observer's reference frame
\begin{equation}\label{eq:kappap}
\kappa'=\frac{B\epsilon}{mc^2\Gamma}\left|\b e_{B}\times\left(\b\eta-\left(1-\frac{\Gamma}{\Gamma+1}\b V\b \eta\right)\Gamma\b V\right)\right|,
\end{equation}
where $\b e_{B}=\b B/B$ is the direction of the magnetic field. Using Eq.~(\ref{eq:el_field}) for the electric field in the force-free magnetosphere one can obtain \citep{bog14}
\begin{equation}\label{eq:vdrift}
\b V=\rho(\b e_{\phi}-(\b e_{\phi}\b e_B)\b e_B),
\end{equation}
where $\rho=|\b r\times \b \Omega|/c=r\sin\theta/R_{lc}$ and $\b e_{\phi}$ is the unit vector along azimuthal direction (direction of rotation). Substitution of Eq.~(\ref{eq:vdrift}) to Eq.~(\ref{eq:kappap}) results in
\begin{equation}\label{eq:kappap1}
\kappa'=\frac{B\epsilon}{mc^2\Gamma}\left|\b e_{B}\times\left(\b\eta_{\perp}-s\b e_{\phi} \right)\right|,
\end{equation}
where $s=\left(1-\frac{\Gamma}{\Gamma+1}\b V\b \eta\right)\Gamma\rho$, and $\eta_{\perp}$ is the perpendicular to magnetic field component of photon direction. Because $s>0$, it is seen from the right panel of Fig.~\ref{fig:vectors} that under Lorentz transformation the perpendicular component of $(\b\eta_{\perp}-s\b e_{\phi})$ increases on the field lines twisted in the direction of rotation and decreases in the opposite case, thus, increasing or decreasing the absorption rate.} 

\begin{figure}. 
	\centering
	
	\mbox{\includegraphics[width=0.5\textwidth]{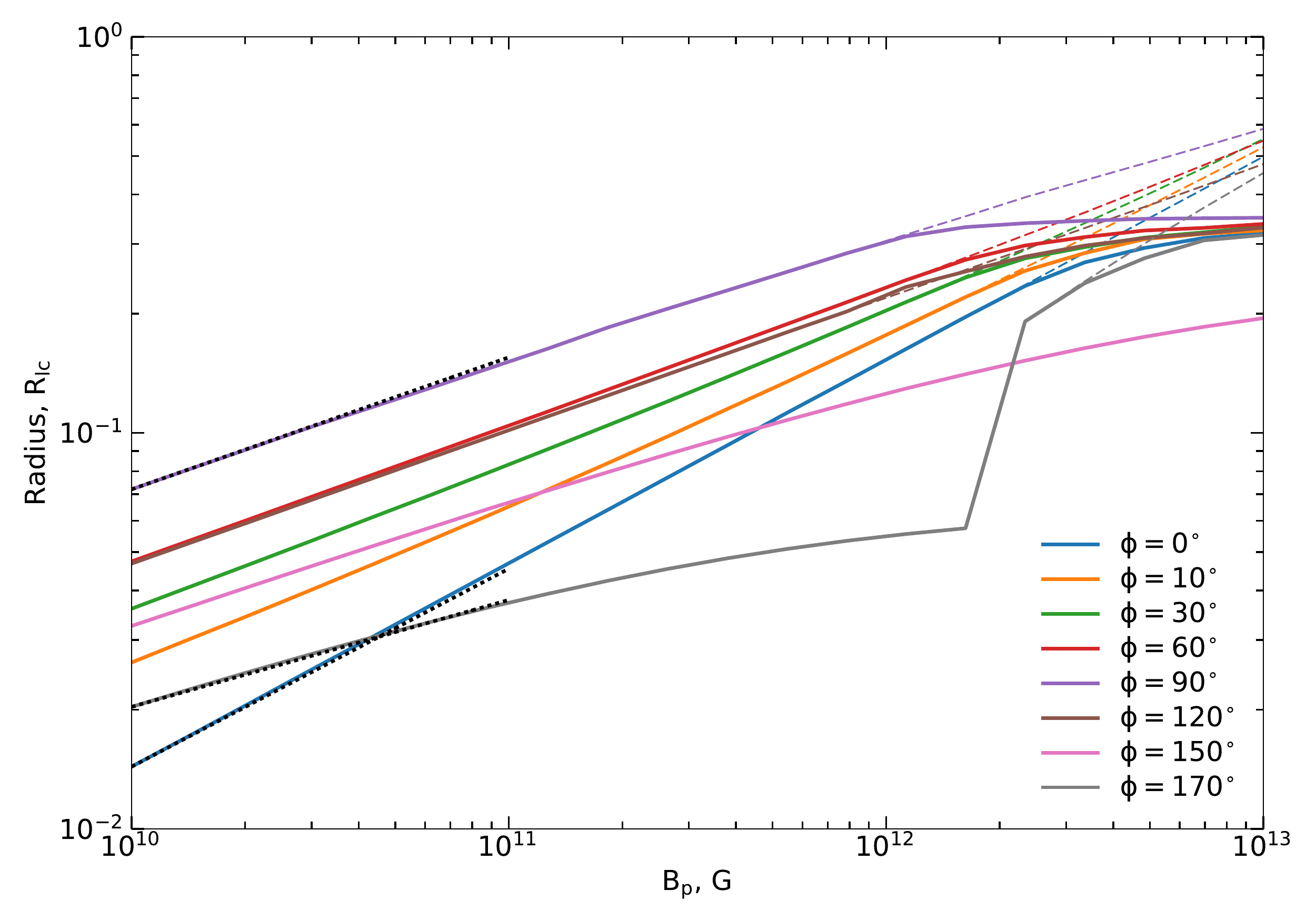}
		\includegraphics[width=0.5\textwidth]{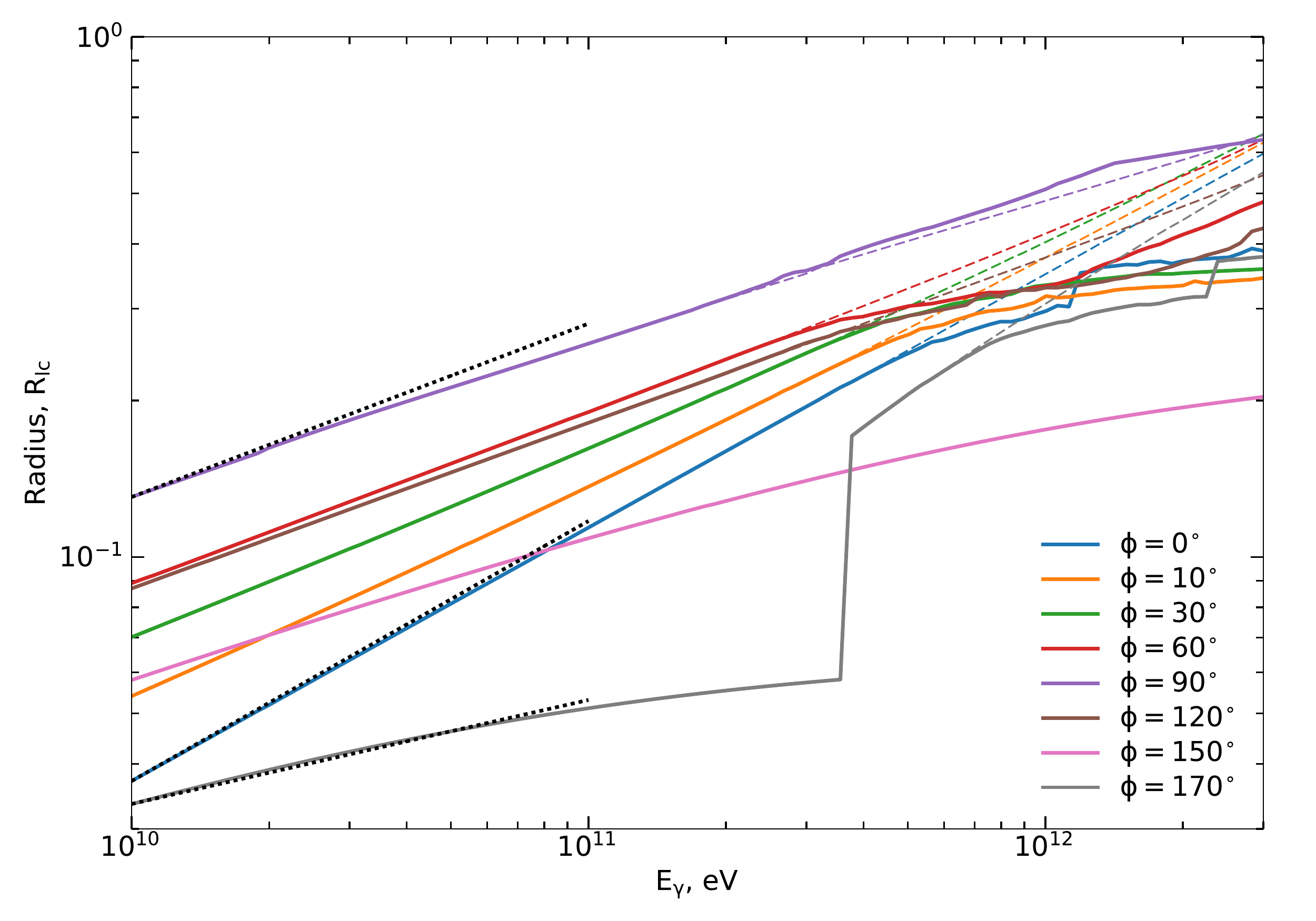}}
	
	\begin{minipage}[b]{\textwidth}
		\caption{Left panel: The radius of the $\tau=1$ surface for the photon with energy $E_\gamma=4\times10^{11}$~eV in equatorial plane for different directions depending on the magnetic field at the pole of the star with radius $R_*=15$~km. The solid and dashed lines represent results for force-free and dipole magnetosphere, respectively. The dotted lines are the guide to the eye power-law lines $\sim B_p^\alpha$ with $\alpha=1/2,1/3,1/4$ for the lines fitting the lines along $\phi=0^{\circ}, 90^{\circ}, 170^{\circ}$. Right panel: The same as left panel but depending on the photon energy for the fixed magnetic field at the pole $B_p=2.1\times10^{12}$~G. The dotted lines are the guide to the eye power-law lines $\sim E_{\gamma}^\alpha$ with $\alpha=1/2,1/3,1/5$ for the lines fitting the lines along $\phi=0^{\circ}, 90^{\circ}, 170^{\circ}$.}\label{tau_bp}
	\end{minipage}
\end{figure}

Although we consider the escape of TeV photons from the Crab pulsar magnetosphere, it is instructive to understand how the results are changed in the case of different magnetic fields and photon energies. The left panel of Fig.~\ref{tau_bp} shows that in the dipole magnetic field the radius to the $\tau=1$ surface changes approximately as a power-law $\sim B_p^{\alpha}$ with $\alpha=1/4-1/2$ depending on the angle $\phi$ of the direction relative to the magnetic moment. The dependence quickly changes from $\alpha \approx 1/2$ at $\phi=0^{\circ}$ to $\alpha \approx 1/3$ at $\phi=10^{\circ}$, and gradually softens up to $\phi=150^{\circ}$, where it starts to drop to $\alpha=1/4$. Using the radius values at different angles for particular magnetic field one can roughly recover the $\tau=1$ surface in equatorial plane similar to the one presented in the right panel of Fig.~\ref{fig:proj_z0}. For the force-free magnetosphere the dependence becomes very slow at the magnetic field strengths $B_p>10^{12}$~G. In this regard, the uncertainties for the mass and the radius of the star in the evaluation of the magnetic field, which scales as $B\sim M^{1/2} R_*^{-2}$, performed in Section~\ref{sec:2} introduce very weak uncertainties in the position of the $\tau=1$ surface. The fast increase of the radius for $\phi=170^{\circ}$ at large magnetic field strength occurs because of the slight "rotation" of the $\tau=1$ surface close to the pole with strength of the magnetic field. The $\tau=1$ surface demonstrates similar behavior for the dependence on photon energy at the fixed magnetic field as it is shown in the right panel of Fig.~\ref{tau_bp}. It is seen that this dependence is slightly softer compared to the dependence on the magnetic field. For example, at $\phi=170^{\circ}$ the dependence on the magnetic field is $\sim B_p^{1/4}$ (the fitting dotted line), whereas the dependence on the photon energy is $\sim E_{\gamma}^{1/5}$.

Note that in  the case of the Crab pulsar, the equatorial plane does not give an information about the observed emission. Indeed, electrons moving in the equatorial plane do not produce radiation detectable on Earth since we observe the Crab pulsar at the angle $57^{\circ}$ relative to the axis of rotation \citep{Hester1995}. Therefore, one should consider the three-dimensional structure of the $\tau=1$ surface together with the sites producing VHE radiation detectable on Earth.

\subsection{The surface of VHE emission directed towards the Earth}   

Relativistic particles radiate photons in the direction of their motion. We observe radiation from particles the velocity of which in some moment of their motion was pointed toward the Earth. Thus, the calculation procedure is reduced to the search of the points with electron velocities constituting the observation angle with the axis of rotation.

\begin{figure}. 
 \centering

	\mbox{\includegraphics[width=0.5\textwidth]{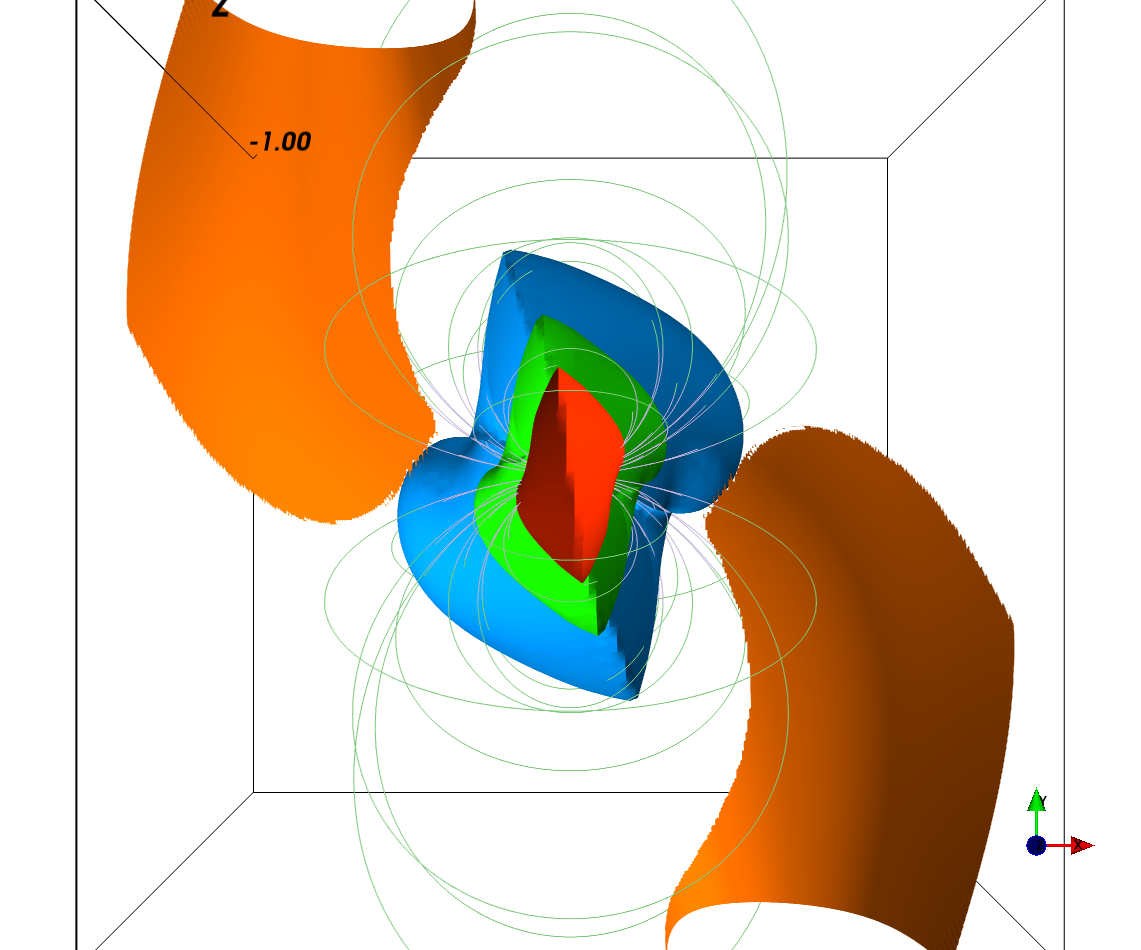}
	\includegraphics[width=0.5\textwidth]{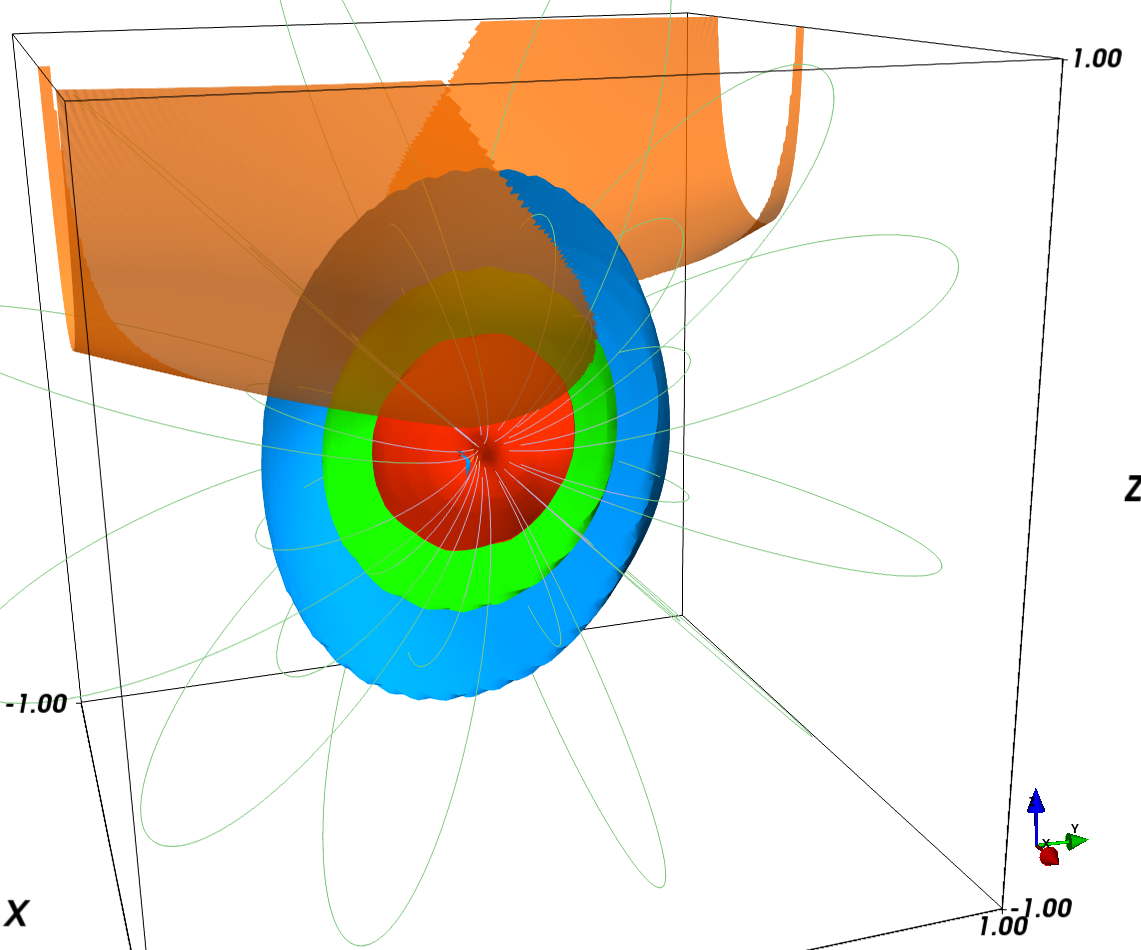}}
	
	\begin{minipage}[b]{\textwidth}
		\caption{Left panel: The surface where photons are emitted towards  the Earth (orange colored regions) (emission at an angle of $33^{\circ}$ with respect to the pulsar equatorial plane), and surfaces $\tau=1$ for $E_{\gamma}=10^{11}$~eV (red), $E_{\gamma}=4\times 10^{11}$~eV (green) and $E_{\gamma}=1.5\times 10^{12}$~eV (blue) photons in the case of the dipole magnetic field with $B_p=2.1\times10^{12}$~G at the poles of the pulsar with radius $R_*=15$~km. View from the top in the direction opposite to $z$-axis. Right panel: The same surfaces in a different projection (axes shown at bottom right). Thin lines present the field lines of the dipole magnetic field.}\label{dipole3d}
	\end{minipage}
\end{figure}

First, we consider the dipole magnetic field.  The view on this surface from above the $z$-axis is shown in Fig.~\ref{dipole3d}. For clarity we present this surface in another projection on the right panel of this figure.
We show only a part of the surface of the photon emission which is located in the region of open field lines. It is seen that everywhere on this surface $\tau < 1$, i.e. this surface is located further away from the pulsar than the surface $\tau=1$  for all considered energies. This can be seen from the projected view in Fig.~\ref{fig:surf33}. The smallest distance from the pulsar to the surface of radiation in the region of open field lines is $\approx 0.4\,R_{\rm{lc}}$ (see the intersection of dashed red and violet lines in Fig.~\ref{fig:surf33}). 

\begin{figure}. 
 \centering

	\mbox{\includegraphics[width=0.5\textwidth]{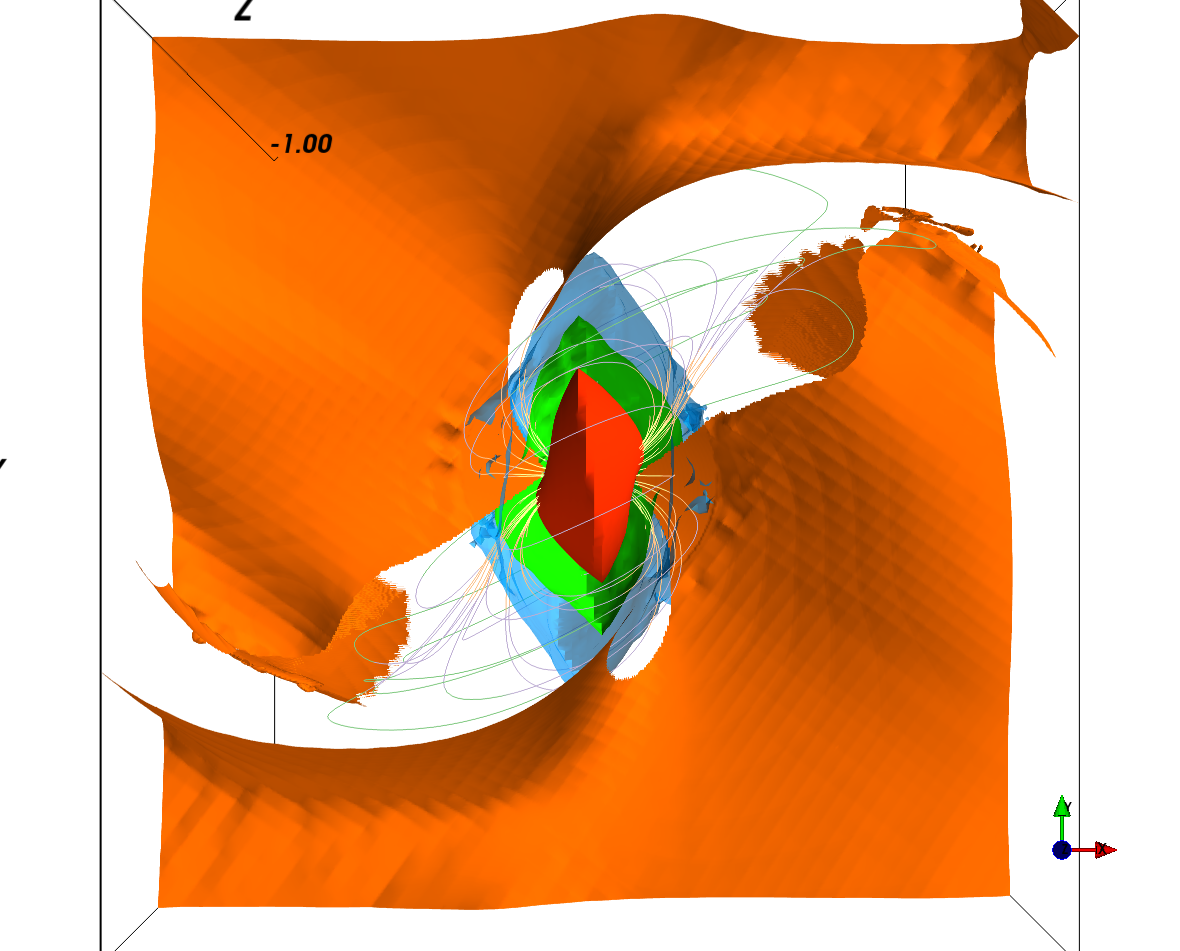}
	\includegraphics[width=0.5\textwidth]{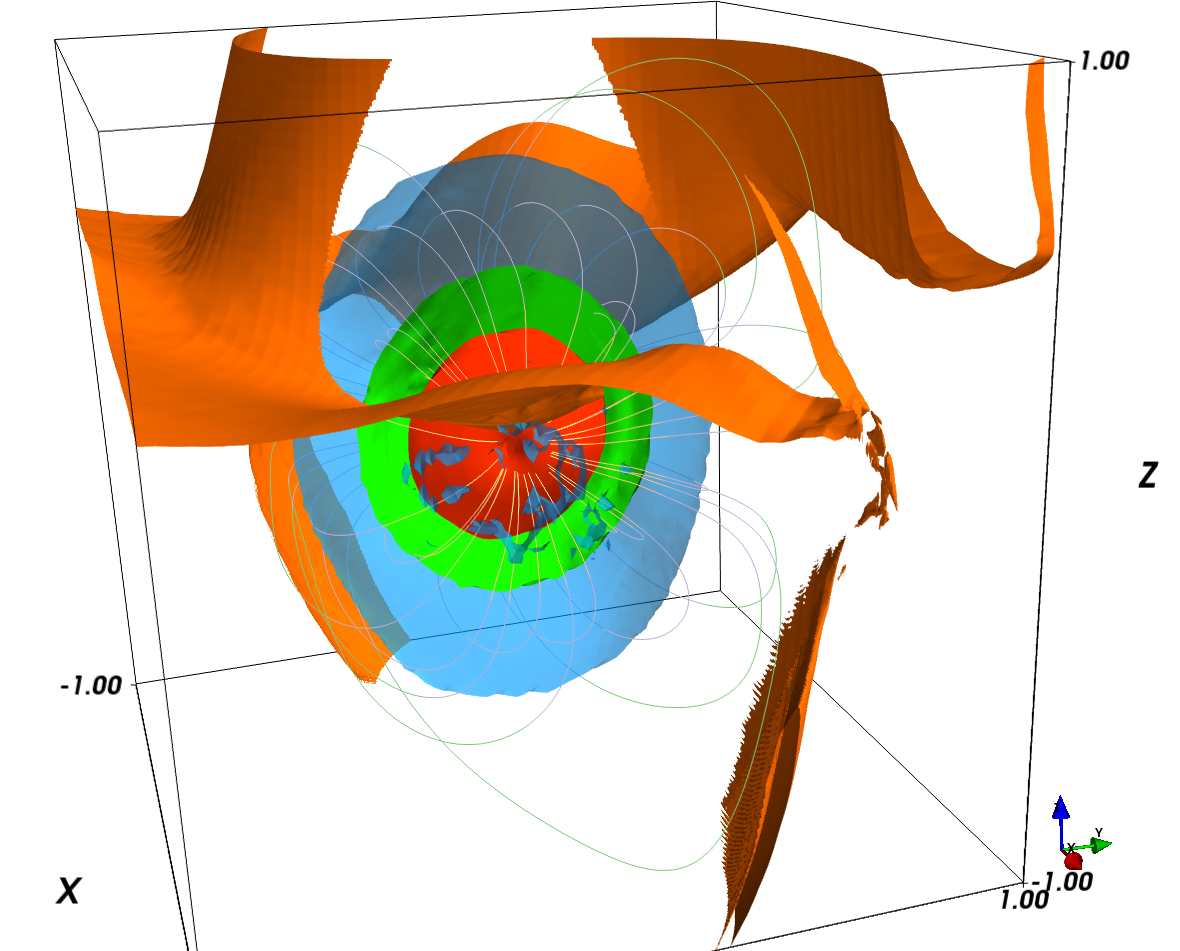}}
	
	\begin{minipage}[b]{\textwidth}
		\caption{The same as Fig.~\ref{dipole3d} but for the force-free magnetosphere.}\label{ffree3d}
	\end{minipage}
\end{figure}

\begin{figure}
	\centering
	\includegraphics[width=0.5\textwidth]{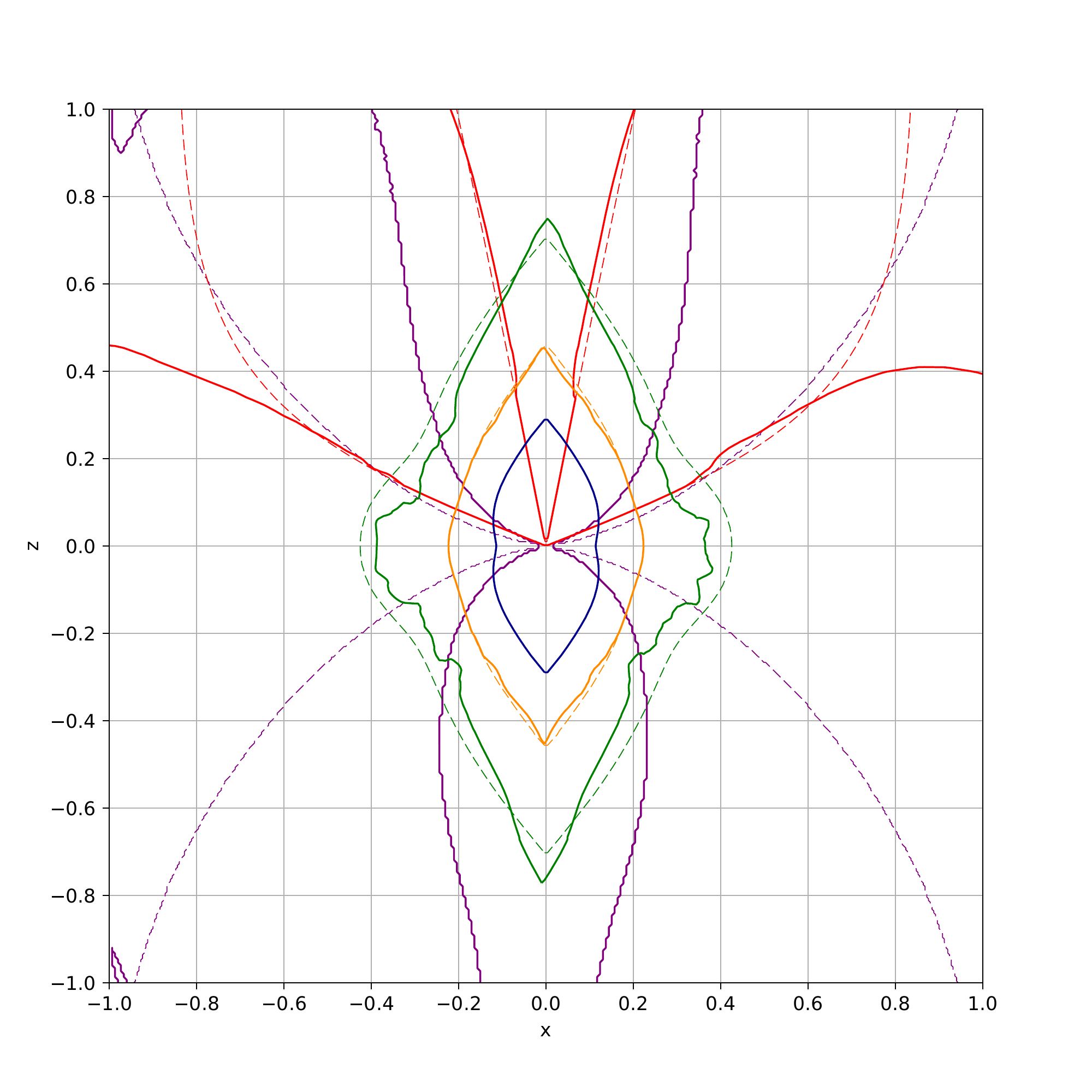}
	
	\caption{Cut of Figs.~\ref{dipole3d} and \ref{ffree3d} in the plane of rotation axis and magnetic dipole (XZ plane at $y=0$). The closed lines in the center are projections of the surfaces $\tau=1$ for $E_{\gamma}=10^{11}$ eV (blue), $E_{\gamma}=4\times10^{11}$~eV (orange), and $E_{\gamma}=1.5\times 10^{12}$~eV (green) photon energies. The red line is the surface emitting photons at an angle of $33^{\circ}$ with respect to the equator.
    The violet lines are the surfaces of closed field lines. The dashed lines are related to the dipole magnetic field and  Fig.~\ref{dipole3d}, whereas the solid lines are for the force-free magnetosphere and Fig.~\ref{ffree3d}.}\label{fig:surf33}	
\end{figure}

At small distances, the surfaces emitting photons in the direction of the Earth coincide for the dipole and force-free magnetospheres since the magnetic fields do not differ there. However, at large distances these surfaces are completely different as seen in Figs.~\ref{dipole3d} and \ref{ffree3d}. As shown in Fig.~\ref{fig:surf33}, the opening angle of the open field line region is significantly larger in the case of the force-free magnetosphere. Therefore, in the force-free magnetosphere the restriction on the emission surface is more relaxed, and the smallest distance from the pulsar to the surface of radiation in the region of open field lines is $\approx 0.09\,R_{\rm{lc}}$. Thus, in the case of the force-free magnetosphere the consideration  of gamma-ray absorption plays an important role in the restriction of the emission region compared to the dipole magnetic field, where the limits are of pure geometrical nature.
The restrictions due to absorption are valid even for photons with energy $E_{\gamma}=10^{11}$~eV. However, we should note that for sufficiently large observation angles relative to the pulsar equator the geometrical factors can be more restrictive compared to the absorption.

\section{Discussion}
\label{sec:4}
The study of transparency of the realistic model of the pulsar magnetosphere for gamma-rays performed in this paper shows that the gamma-ray absorption can be small even for photons of energies $E_{\gamma}=1.5$~TeV. The minimum distance from the pulsar where these photons can escape from the magnetosphere appears close to $r_{\rm{min}}=0.1\,R_{\rm{lc}}$. Obviously, this distance depends on the field line where the photon has been produced. Along the last closed field line twisted in the direction of rotation, this distance reaches $0.3\,R_{\rm{lc}}$. The rest of the magnetosphere in the region of open field lines appears transparent for gamma-rays with energy $1.5$~TeV and below. The magnetosphere appears more opaque for photons traveling along field lines twisted in the direction of rotation than in the opposite direction.

{ The force-free magnetosphere appears more transparent for VHE gamma-rays above $1.5$~TeV than the magnetosphere of the dipole magnetic field. This phenomenon has the same reason as the asymmetry of the absorption for the photons traveling along field lines with different twist. Namely, the aberration in the electric field increases or decreases the perpendicular to the magnetic field component of the photon direction on the field lines twisted along or opposite to the rotation, respectively. The field lines of the force-free magnetosphere are twisted preferably in the direction opposite to the rotation reducing absorption, whereas the twisting direction of the dipole field lines can be both along and opposite to the rotation.}

The consideration of the surface emitting photons towards observer in the region of open field lines gives further restrictions on the VHE emission region. We discuss the limitations for the observation angle $57^{\circ}$ relative to the axis of rotation relevant to the Crab pulsar. For this angle, in the case of dipole magnetic field the emission surface in the region of open field lines is located outside of the region of strong magnetic absorption and the detectable VHE photons should be emitted from a distance larger than about $0.4\,R_{\rm{lc}}$. In the case of the force free magnetic field the emission surface is noticeably larger due to wide open field line region and the absorption remains the main restriction on the minimum distance of the emission. Apparently, this result depends strongly on the implied magnetic field geometry. We have performed our calculations only for the case of orthogonal rotation. Obviously, different inclination angles will yield different results. In this sense, magnetic absorption yields more reliable and general restrictions, because deviation from orthogonal rotation modifies only weakly the absorption of photons in the magnetic field. The same is valid in relation to the value of the magnetic field. Variation of the magnetic field results in small changes in the position of the $\tau=1$ surface near the stellar surface. 

{ Let us discuss the contribution of $\gamma\gamma$ absorption of VHE gamma rays on the thermal radiation of the neutron star to the overall absorption. For its evaluation we approximated the thermal emission of the Crab neutron star by an isotropic photon field with a blackbody spectrum of a temperature $T=1.9\times10^6$~K and total luminosity $10^{34}\,\rm{erg}\,\rm{s}^{-1}$ \citep{Becker1997}.
Because the absorption significantly changes with the distance from the star as $\sim 1/r^{2}$, we calculated for the $\gamma\gamma$ absorption the $\tau=1$ surface similar to the one depicted in Fig.~\ref{fig2}. The position of the $\tau=1$ surface relative to the $\gamma\gamma$ absorption changes with energy approximately as $E_{\gamma}^{-0.9}$. The maximum distance of the surface in equatorial plane is located at $7\times10^{-3}$, $2.5\times10^{-3}$ $8\times10^{-4}\,R_{\rm{lc}}$ for  $E_{\gamma}=10^{11},\, 4\times10^{11}, \, 1.5\times 10^{12}$~eV, which is smaller than the radius of the neutron star $R_*=9.5\times10^{-3}\,R_{\rm{lc}}$.
It means that the VHE gamma rays are not absorbed due to $\gamma\gamma$ interaction with the considered photon field. Only at  $E_{\gamma}\approx3\times10^{9}$~eV, where the maximum absorption occurs, the $\tau=1$ surface is located at the distance $0.038\,R_{\rm{lc}}$, which is of the order of several star radii. The left panel of Fig.~\ref{tau_bp} shows that in some directions at this energy the magnetic absorption could be slightly lower than $\gamma\gamma$ absorption.}

In summary, the absorption of gamma-rays in the pulsar magnetosphere does not impose strong limitations on the site of the VHE production. Photons with energy $400$~GeV  can be produced at the distance $0.06\,R_{\rm{lc}}$ from the pulsar. This information,
combined with specific models for electron acceleration in the pulsar magnetosphere, can give much stronger limitations. For example, close to the region of the last closed field lines (important for outer gap model) the potential emission distances increase almost an order of magnitude. Therefore, the result of our work will be especially helpful in the context of specific models for the electron acceleration in the pulsar magnetosphere.

\section*{Acknowledgments}
The present work was supported by Russian scientific fund, project N 16-12-10443. The numerical simulations of the force-free magnetosphere were performed at the Research Center for Astronomy and Applied Mathematics of the Academy of Athens using the resources of the NRNU MEPhI High-Performance Computing Center.

\appendix
\label{sec:app}
\section{$\gamma B$ photon conversion into \lowercase{$e^{\pm}$} pairs}
\label{sec:aconv}
We restrict ourselves to the case of a force-free magnetosphere where the condition $\b E \cdot \b B=0$ holds everywhere. This allows us to find at any point a reference frame in which the electric field $\b E$ disappears, thus reducing the calculations of magnetic pair conversion to the case of a pure magnetic field. The total probability of pair production in a pure magnetic field per unit length (attenuation coefficient) can be expressed in the form of \cite{Tsai1974,Urrutia1978} \footnote{For the alternative form for the attenuation coefficient see \cite{Nikishov1967}}
\begin{equation}\label{eq:att_length}
\mathcal{R}=\frac{\sqrt{3}}{9\pi}\frac{\alpha}{\lambdabar_C}\frac{1}{\epsilon}\int\limits_{0}^{1}
{\rm d}v\,\left( \frac{9-v^2}{1-v^2}\right) K_{2/3}\left(\frac{8}{3(1-v^2)\kappa}\right),
\end{equation}
where $\alpha=e^2/\hbar c$ is the fine-structure constant, $\lambdabar_C=\hbar/mc$ is the reduced Compton wavelength, $\epsilon=\hbar\omega/mc^2$ is the photon energy in units of the electron rest mass, $\kappa=\widetilde{B}\,\epsilon\sin\theta$, $\widetilde{B}=B/B_{cr}$ is the magnetic field strength normalized to the critical magnetic field $B_{cr}=m^2c^3/e\hbar$, $\theta$ is the angle between the photon momentum and the magnetic field, and $K_{2/3}(x)$ is the modified Bessel function of the second kind.

Note that Eq.~(\ref{eq:att_length}) is an approximation valid in the limit of high-energy photons $\epsilon\sin\theta\gg1$ and weak fields $\widetilde{B}\ll1$. A significantly more complicated expression valid for arbitrary magnetic fields can be found in \cite{Tsai1974}. However, the $\widetilde{B}\gtrsim 1$ regime is not relevant for the regular pulsars considered in the current work whose magnetic field strengths are usually an order of magnitude smaller than the critical strength $B_{cr}\approx4.4 \times 10^{13}$~G.

The requirement of the high-energy photon limit $\epsilon\sin\theta\gg1$ is more subtle, because it could break down in the small region close to the emission point where the photon is emitted almost along the magnetic field with $\theta\ll 1$. As shown by \cite{Daugherty1983}, the approximation given by Eq.~(\ref{eq:att_length}) is several orders of  magnitude higher than the exact expression calculated near the threshold. Nevertheless,
in the near threshold region the attenuation coefficient $\mathcal{R}$ for both expressions rises fast while the difference quickly reduces with the increase of $\epsilon\sin\theta$. We are interested in the optical depth, which is the integral of the attenuation coefficient $\mathcal{R}$, therefore the region close to the threshold will not play any role, as long as the main contribution to the optical depth is given by the region $\epsilon\sin\theta\gg1$. Since we study the absorption of high-energy photons, which occurs at sufficiently large $\theta$, we can neglect the near threshold effects.

Having the expression for the attenuation coefficient in the reference frame where the electric field disappears, one can calculate the optical depth using the following procedure. Assuming that $E<B$ and $\b E \cdot \b B=0$, one can consider the absorption in the reference frame moving relative to the stationary observer's reference frame with the velocity $\b v/c\equiv\b V=(\b E\times \b B)/B^2$. It is then straightforward to see that the electric field $\b E'$ in this reference frame disappears, and that the magnetic field becomes $\b B'=\b B/\Gamma$, preserving its direction and reducing its magnitude by a factor of $\Gamma$, where $\Gamma=1/\sqrt{1-(E/B)^2}$. Correspondingly, the direction $\b \eta'$ and energy $\epsilon'$ of a photon in the new reference frame are
\begin{eqnarray}\label{eq:abber}
\hspace{0cm}
\b \eta'=\delta\left(\b \eta+\left(\frac{\Gamma}{\Gamma+1}\b V\b \eta-1\right)\Gamma\b V\right), \quad \epsilon'=\frac{\epsilon}{\delta},
\end{eqnarray}
where $\delta=1/\Gamma(1-\b V \b \eta)$ is the Doppler factor. Substituting $\b B'$, $\epsilon'$, $\b \eta'$ into Eq.~(\ref{eq:att_length}) we obtain the attenuation coefficient $\mathcal{R}'$ in the moving reference frame. The attenuation coefficient in the observer's reference frame is simply equal to $\mathcal{R}=\mathcal{R}'/\delta$.
Since $\mathcal{R}'\sim 1/\epsilon'=\delta/\epsilon$, for calculation of $\mathcal{R}$ in the observer's reference frame we should effectively use Eq.~(\ref{eq:att_length}) with $\kappa'$ instead of $\kappa$.

\section{Procedure of the calculation of the magnetic absorption with aberration}
\label{sec:acalc}
We start the calculation of the optical depth with the determination of the direction of photon emission. The high-energy photons are emitted by ultra-relativistic electrons and positrons ($\gamma \gtrsim 10^5$) in a narrow cone with an angle of $1/\gamma$ around their direction irrespective to the radiation mechanism. Therefore, we can assume that photons move almost exactly along the direction of the parent electrons. The motion of the charged particle in the crossed electric and magnetic fields can be approximated as a motion along the magnetic field with a simultaneous electric drift perpendicular to it:
\begin{equation}\label{eq:el_dir}
\b \beta=\frac{\b E\times \b B}{B^2}\pm f\frac{\b B}{B},
\end{equation}
where the sign $\pm$ corresponds  to the two  opposite directions along a magnetic field line \citep[see e.g.][]{Bai2010a}. The  factor $f=\sqrt{1-E^2/B^2}$ is defined from the condition of ultra-relativistic motion $\beta\approx1$. 
Note that the assumption that electrons move strictly along magnetic field lines relies on the smallness of the Larmor radius in comparison with the curvature radius of the field line. In fact, we use a 0-th approximation, in which the Larmor radius is assumed to be equal to zero. The inclusion of higher order terms leads to the inertial drift of the electrons \citep{Kelner2015}.

The direction of motion of the photon given by Eq.~(\ref{eq:el_dir}) is not changed in the observer's reference frame. Thus we calculate the optical depth as
\begin{equation}\label{eq:tau}
\tau=\int\limits_{0}^{s}\,\mathcal{R}ds'.
\end{equation} 
The integration is performed along the path $\b r(s)=\b r_0+s\,\b \eta$, where $\b r_0$ is the emission point, and $\b \eta=\b \beta$ is the photon direction. The attenuation coefficient $\mathcal{R}=\mathcal{R}(\b r(s))$ varies through its dependence on the magnetic $\b B(\b r(s),t)$ and electric $\b E(\b r(s),t)$ fields, where $t=t_0+s/c$. The function $\mathcal{R}=\mathcal{R}(\b r(s))$ depends on the coordinates and time due to the rotation of the magnetosphere.
The co-rotation of the plasma in the force-free limit implies a relation between the electric and magnetic fields, namely
\begin{equation}\label{eq:el_field}
\b E=-\frac{\b\Omega\times\b r}{c}\times \b B,
\end{equation} 
where $\b \Omega$ is the pulsar angular velocity.

The assumption of a stationary pattern of magnetic field $\b B_0(\b r)$ co-rotating with the star reduces the problem to the rotation transformations
\begin{equation}\label{eq:rot}
\b B(\b r, t)=\mathcal{A}\b B_0(\mathcal{A}^{-1}\b r),
\end{equation}
where $\mathcal{A}=\mathcal{A}(t)$ is rotation matrix. In particular, setting the reference frame so that $\b \Omega$ is along $z$-axis, the rotation matrix has the form
\begin{equation}
\mathcal{A}(t)=\mathcal{M}_z(\Omega t+\phi_0)\mathcal{M}_y(\alpha),
\end{equation}
where $\mathcal{M}_{\bf \eta}(\theta)$ rotates a vector around axis ${\bf \eta}$ by an angle $\theta$, $\phi_0$ is the initial pulsar phase, and $\alpha$ is the magnetic inclination angle. Specifying the magnetic field distribution $\b B_0(\b r)$ (dipole or force-free magnetic field in the current work) at time $t=0$ determines the calculation of the optical depth $\tau$ given by Eq.~(\ref{eq:tau}) through the sequential applications of Eqs.~(\ref{eq:rot}), (\ref{eq:el_field}), (\ref{eq:el_dir}), (\ref{eq:abber}), (\ref{eq:att_length}).


\label{lastpage}

\end{document}